\documentstyle[preprint,aps,epsf]{revtex}
\begin{document}
\title{\Large\bf Characteristic Crossing Points in Specific Heat Curves
of Correlated Systems}

\author{D.~Vollhardt\thanks{New permanent address after
September 15, 1996:
Theoretische Physik III, Institut f\"{u}r Physik, Universit\"{a}t
Augsburg, 86135 Augsburg, Germany; e-mail: vollha@physik.uni-augsburg.de} }
\address{  Institut f\"{u}r 
Theoretische Physik C, Technische Hochschule Aachen,\\
D-52056 Aachen, Germany}

\date{ \today}
\maketitle

\begin{abstract}
Attention is drawn to the observation that in many correlated
systems (e.g. $^3$He, heavy fermion systems and Hubbard models)
the specific heat curves, when plotted for different values of 
some thermodynamic
variable (e.g. pressure, magnetic field, interaction), cross
almost precisely at one or two characteristic temperatures.
A quantitative explanation of this phenomenon, based on
the temperature dependence of the associated generalized susceptibilities,
is presented.

\vspace{12pt}
\noindent
PACS: 67.55.Cx, 71.27.+a,71.28.+d
\end{abstract}

\vspace{12pt}
In 1959 Brewer et al. \cite{1} noticed that the specific heat curves
$C(T,P)$ of liquid $^3$He measured at different pressures $P$ all 
intersect at a temperature $T_+ \simeq 0.16 K$, and that $C(T,P)$ 
{\it increases} with increasing pressure below $T_+$. Greywall's 
high-precision measurements of the specific 
heat at constant volume
$V$ provided striking evidence for the sharpness
of the crossing point  at $T_+ \simeq 160$mK (see Fig.~1a)\cite{2}.
At this temperature the specific heat is obviously independent
of volume or pressure. It was unclear, however, whether special
significance should be attached to this finding \cite{2,4}. Recently, 
Georges and Krauth \cite{5} observed the same conspicuous crossing 
phenomenon in quite a different system,
namely in the paramagnetic phase of the Hubbard model, the simplest 
model of correlated electrons, in infinite dimensions.
For small
to intermediate values of the local interaction $U$ the specific
heat curves $C(T,U)$, calculated by iterated 
perturbation theory, were found to intersect almost at the same 
temperature $T_+ \simeq 0.59t^*$, where $t^*$ is the scaled
hopping amplitude of the electrons (Fig. 2). 
Clearly the existence of
these peculiar points of intersection  calls for an explanation.

In this Letter we illustrate that crossing points 
 such as the ones described above can actually
be observed in the specific heat of many 
correlated systems. Furthermore, we explain the
origin of this phenomenon.

To be able to discuss the problem in a sufficiently general frame
work  we define a general free energy 
$\Phi (T, X)$, where $X$ can 
be any thermodynamic variable, e.g. pressure $(P)$, magnetic field
$(B)$, on-site interaction $(U)$. The
conjugate variable associated with $X$ is
$\xi (T,X) = - \partial \Phi (T,X)/\partial X$.
Pairs of variables $(X, \xi)$ are, for example,
$(P, -V), (B, M)$ and $(U, -D)$, with $M$ as the magnetic moment, and $D$
as the number of doubly occupied sites in
Hubbard models. With the entropy $S(T,X) = - \partial \Phi/\partial T$ 
one obtains the Maxwell relation
\begin{equation}
\frac{\partial S (T,X)}{\partial X} = \frac{\partial \xi (T,X)}{\partial T} .
\label{Gl2}
\end{equation}

A search of the literature reveals that there exists quite a number
of systems, both in theory and experiment, where the specific heat 
curves $C(T,X)
= T \partial S(T,X)/\partial T$ versus $T$ when plotted for different,
not too large
values of $X$ intersect at one or even two well-defined, non-zero
temperatures. Apart from normal-liquid $^3$He it can be observed
in heavy fermion systems with and without Fermi liquid behavior,
 for  example in CeAl$_3$ \cite{6} (Fig.~3a)
 and UBe$_{13}$ \cite{7} upon change of $P$, in
UPt$_{3-x}$Pd$_x$ \cite{8}
and CePt$_3$Si$_{1-x}$Ge$_x$ \cite{8a}
as $x$ is varied, and in CeCu$_{6-x}$Au$_x$ 
$(x = 0.3, 0.5$) when either $P$ \cite{9} or $B$  \cite{10}(Fig.~3b)
is varied. It is also found in Eu$_{0.5}$Sr$_{0.5}$As$_3$, a semi-metal
with competing interactions, upon change of B \cite{11}. In particular,
all theoretical models of Fermi- and Luttinger liquids 
investigated beyond the low-temperature regime show this features:
the one-dimensional $(d = 1)$ Hubbard model 
in a magnetic field \cite{13}, the 
$1/r$-Hubbard in $d = 1$ in the metallic phase
when the interaction $U$ is changed
\cite{14}, and the 
Hubbard model in $d = \infty$ discussed above \cite{5}.

To explain the origin of the crossing points we separate the problem into
two questions: (i) Why do specific heat curves cross at all?, and (ii) 
How wide is the region where the curves cross? Turning to the first question, we note that any
crossing of specific heat curves $C(T,X)$ implies
\begin{equation}
\frac{\partial C(T,X)}{\partial X} \bigg|_{T_+ (X)} = 
T_+ (X) \frac{\partial^2 \xi (T,X)}{\partial T^2} \bigg|_{T_+ (X)} = 0.
\label{Gl3}
\end{equation}
Thus crossing occurs where $\xi (T,X)$ versus $T$ has a turning
point. In general the crossing temperature $T_+(X)$ still
depends on $X$. Only if $T_+$ is {\it independent}
of $X$ for some range of $X$-values do the curves intersect at 
one point. Crossing of specific heat curves may be inferred from a
sum rule for the change of the entropy $S(T,X)$ with 
respect to $X$ in the limit 
$T \to \infty$
\begin{equation}
\eta_X^{}  = k_B^{-1} \lim_{T \to \infty} \frac{\partial S (T,X)}
{\partial \ln X} = \frac{X}{k_B} \int_0^\infty
\frac{dT'}{T'} \frac{\partial C(T',X)}{\partial X}.
\label{Gl4}
\end{equation}
1. \underline{Lattice models ($X \equiv U$):}
Eq.~(\ref{Gl4}) implies $\eta_U^{} = 0$ for any kind of Hubbard model
since $S$ approaches a constant for $T \to \infty$.
(At high temperatures, $T \gg U, C(T,U)
\propto U/T$, i.e. $\partial C/\partial U > 0)$.
 Hence
$\partial C/\partial U$ must have positive and negative contributions,
i.~e. the specific heat curves
must cross at one or more temperatures, for the integral
to vanish identically. We note that this 
is a genuine {\it correlation} effect
originating from the existence of $U^2$ and higher terms in an
expansion of the internal energy $E(T,U)$, and hence of
$C(T,U) = \partial E/\partial T$, in powers of $U$.

\noindent
2. \underline{Continuum systems ($X \equiv P)$:}
Here eq.~(\ref{Gl4}) implies 
$\eta_P^{} = - 1$ since $S$ 
approaches the ideal-gas value for $T \to \infty$.
Apparently $\partial C/\partial P < 0$ at most (especially
{\it high}) temperatures \cite{15}. Now, in the Fermi liquid phase
of $^3$He \cite{1,2} and in the paramagnetic phase of
the Hubbard model at low temperatures \cite{5,13} the entropy
is known to {\it increase}
with $X (\equiv P,U)$, i.e. $\partial C/\partial X > 0$. Hence, at sufficiently 
low $T$ when $S(T,X) = C(T,X) =
\gamma T$, one has $d \gamma /dX > 0$. This may
be attributed to the excitation of low-energy (spin) degrees of
freedom in the correlated system \cite{16}. Eq. (3) then implies that
the specific heat curves of liquid $^3$He cross
once (Fig.~1a) and in the Hubbard model twice
(Fig.~2). These two systems only consist of a single 
species of particles and there exists {\it one} characteristic
low-temperature scale -- the crossing temperature $T_+ (X)$.
(In fact, in $^3$He  $T_+$ practically coincides 
with the temperature above which
Fermi liquid theory breaks down \cite{17}). 
By contrast, heavy fermion systems
are basically two-component systems consisting of conduction and
localized electrons which may hybridize. The strength of the 
hybridization is determined by an amplitude $V_{hyb} (P)$
which increases with pressure. By hybridizing, the electrons 
may gain an energy $k_B T_K$, where $T_K (P) \propto \exp 
(- \mbox{const}/V^2_{hyb})$ 
is a different (``Kondo'') low-energy scale. 
Below $T_K$ the linear specific heat coefficient is given by 
$\gamma (P) \propto 1/T_K$, where now $d \gamma/dP < 0$, and hence the
specific heat {\it decreases} with pressure.
This implies that the specific heat curves will cross {\it twice}:
at $T_+$, below which the  low-energy (spin) excitations lead
to $\partial C/\partial P > 0$, and again at $T_+^\prime < T_K$
below which $\partial C/\partial P < 0$. This is 
precisely what is seen in several heavy fermion systems, e.g. in CeAl$_3$ 
\cite{6} ($T_+^\prime \simeq 5$K, $T_+ \simeq 17$K; see Fig. 3a)
and UBe$_{13}$ \cite{7} ($T_+^\prime \simeq 2.5$K, $T_+ \simeq 9$K) \cite{18}.

We now turn to the question concerning the
width of the crossing region. A plot of $C(T,X)$ versus $X$
(see e.g. Fig. 1b for $^3$He) shows that
$C(T,X)$ depends only weakly on $X$ even for large values
of $X$. Hence we expand $C(T,X)$ in an (asymptotic) series in $X - X_0$, 
with $X_0$ chosen at convenience,
\begin{eqnarray}
C(T,X) \simeq C(T,X_0) + (X - X_0) T \frac{\partial^2 \xi}{\partial T^2}
\bigg|_{X_0} 
\nonumber \\ 
+ \frac{1}{2} (X- X_0)^2 T \frac{\partial^2}{\partial T^2}
\frac{\partial \xi} {\partial X }\bigg|_{X_0} + \ldots
\label{Gl5}
\end{eqnarray}
where we used eq.~(2). At $T_+ (X_0)$ eq.~(\ref{Gl5}) implies
$C(T_+ , X) \simeq C(T_+ , X_0) \left[ 1 + W_{X_0} (X)\right]$
where the (relative) width of the crossing region, $|W_{X_0} (X)| =
|\Delta_{X_0}^{(1)} (X) + 
\Delta_{X_0}^{(2)} (X) + \ldots |$,
is determined by the numbers
\begin{equation}
\Delta_{X_0}^{(n)} (X)  = \frac{(X- X_0)^{n +1}T_+}
{ (n+1)! C(T_+, X_0)} \; \frac{\partial^2}{\partial T^2} 
\chi^{(n)} (T,X_0) \bigg|_{T_+} ,
\label{Gl7}
\end{equation}
with $\chi^{(n)} (T,X)= \partial^n \xi/ \partial X^n$. 
For $|\Delta_{X_0}^{(n)} (X)| \ll 1$
the  $C(T,X)$ curves will intersect
at a well-defined point. The width is seen to be determined by the
curvature (with respect to $T$) of the linear
($n = 1)$ and  non-linear
($n > 1)$  susceptibilities $\chi^{(n)} (T,X)$ at $T_+$ and $X_0$. 
There are two particularly relevant cases in which the $\Delta^{(n)}$ are
small:

\noindent
(i) \underline{Weak $T$-dependence of $\chi^{(n)}$:}
For $X = P, \xi = - V$ the susceptibility
$\chi^{(1)} = - \partial V/\partial P = \kappa_T V$ 
is essentially the isothermal compressibility of the system. 
For $^3$He and heavy-electron liquids 
we can estimate $\Delta_{P_0}^{(n)} (P)$
from the linear-$T$ behavior of the specific heat which, using 
eq.~(\ref{Gl2}),
implies $\chi^{(n)} (T,P) \simeq \chi^{(n)} (0,P) + \frac{1}{2} 
\gamma^{(n+1)}
(P) T^2$, with $ \gamma^{(n)}(X) = \partial^n \gamma/ \partial X^n$.
Thus we obtain $\Delta ^{(n)}_{P_0} (P) \simeq 
\left[\left( 1- P/P_0 \right)^{n+1}
/(n +1)!\right] \left[ P_0^{n+1} \gamma^{(n+1)} (P_0)/ \gamma (P_0) \right]$. 
For $^3$He \cite{2} (with $P_0 = 15 {\rm bar}, P_0^2 \gamma^{(2)}
(P_0) /\gamma(P_0) \simeq 5 \times 10^{-2})$ we find $|W(P)|
\alt 0.03$
for $0 \leq P \leq 30$ bar, i.e. on the scale of
Fig.~1a the crossing
region is indeed essentially confined to a point.
Hence it is the very weak temperature dependence of
$-\partial V/ \partial P$ of liquid $^3$He at low temperatures,
i.e. the almost  linear pressure dependence of the $\gamma$-factor of the
specific heat, which is the origin of the sharp (but not exact) crossing
of specific heat curves at $T_+ \simeq 160$mK. Similarly, 
for the crossing at $T^\prime_+ \simeq 5$K in CeAl$_3$ \cite{6}
(with $P_0 = 4.8$ kbar, $P_0^2 \gamma^{(2)} (P_0)/ 
\gamma(P_0) \simeq 0.4)$
we obtain  $|W(P)|\alt 0.2$ for $P$ between $0.4 - 8.2$ kbar,
which again agrees with the data (Fig. 3a). 

\noindent
(ii) \underline{Linear $X$-dependence of $\chi^{(1)}$:} If $\xi$ is a
linear function of $X$, i.e. $\xi(T,X) = \chi^{(1)}(T) X$, 
as in linear-response theory, 
the crossing
condition, eq.({\ref{Gl3}), takes
the form $d^2 \chi(T_+)/dT^2 = 0$, where $T_+ = T_+ (X_0 = 0)$. 
This implies that $\Delta_{0}^{(n)}$
vanishes identically for $n \geq 1$. In this case all specific heat curves intersect exactly
at one point. The width $|W_0 (X)|$ becomes finite
only through non-linear terms in 
$\xi(T,X) = \chi^{(1)} (T) X + \frac{1}{3!} \chi^{(3)}(T) X^3 + \ldots$,
where $\chi^{(n)}(T) = \chi^{(n)}(T,X= 0)$. 
The lowest-order contribution to the width is $\Delta_0^{(3)} (X)$
which may be quite small. This explains why 
for not too large magnetic fields 
the specific heat curves $C(T,B)$ 
cross at a well-defined temperature
both in the $d = 1$ Hubbard model at $U = const$ \cite{13}, as
well as in CeCu$_{6-x}$Au$_x$ \cite{10,21} (Fig. 3b). 
The same arguments apply
to $C (T,U)$ of the paramagnetic phase of Hubbard models \cite{5,14}
where we now choose $\xi = \tilde{D}(T,U)
= \frac{1}{4} - D (T,U)$ at half filling such that $\tilde{D} (T,0) = 0$. 
To a good approximation $\tilde{D} (T,U)$
is linear in $U$ for not too large $U$ at all temperatures \cite{22}.
We find $|W(U)|\alt 0.05$
for $U \alt 2.5t^*$ in the $d = \infty$ Hubbard model \cite{23}.~ 

In Hubbard models the intersection
of $C(T,U)/k_B = f (T/t, U/t)$ curves is sharp only at {\it high}
temperatures.  At low
temperatures the generation of low-energy excitations leads to a renormalized
energy scale $t \to t_{\mbox{\footnotesize\em eff}} \ll t$. 
Hence a perturbation expansion of $E(T,U)$ or $C(T,U)$ to second order
in $U$ will
be valid only for a small range of $U$-values, implying a wide crossing region,
$|W(U)|\sim 1$,
at low temperatures.

In summary, we showed that the remarkable crossing of specific heat
curves $C(T,X)$ vs. $T$ for different thermodynamic variables $X$, first 
observed in $^3$He \cite{2} and the Hubbard model \cite{5}, is not
accidental but can be found in many correlated systems.
The width of the crossing
region is found to give explicit information about the 
temperature dependence of the generalized
susceptibilities associated with $X$.
A related observation, that for Hubbard models the value of
the specific heat at the crossing point is almost universal,
will be discussed elsewhere \cite{24}.

I am grateful to M. Kollar, G. Baskaran, D. S. Greywall,
H. Keiter,
W. Metzner, E. M\"{u}ller-Hartmann, H. R. Ott, 
H. G. Schlager, F. Steglich and G. R. Stewart for useful 
discussions, and to N. Bl\"{u}mer, N. Chandra,
M. Kollar, R. R\"{o}mer and J. Schlipf
for helping with  the figures. I also thank
R. A. Fisher, F. Gebhard, G. Georges, A. Kl\"{u}mper,
H. v. L\"{o}hneysen and G. R. Stewart for kindly providing
me with data and figures of the specific heat of various
correlated systems. 
This research was supported in part by the SFB 341 of the
Deutsche Forschungsgemeinschaft, and the National Science Foundation
under Grant No. PHY94-07194.

\begin{figure}
\epsfysize16cm
\centerline{\epsfbox{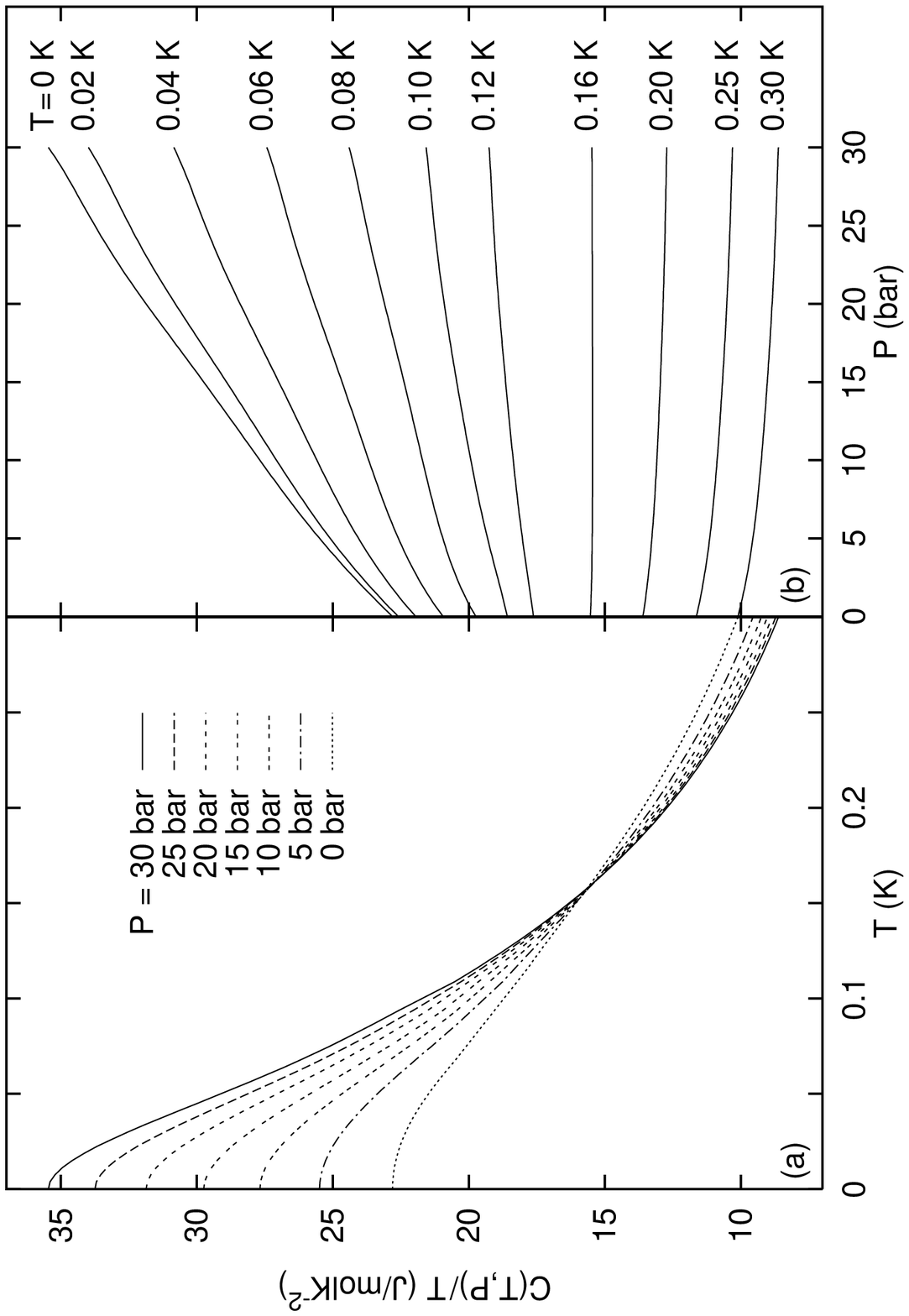}}
\caption{Specific heat
 $C (T,P)$ of $^3$He \protect\cite{2}: (a) $C/T$ vs. $T$, (b) $C/T$ vs. $P$.}
\end{figure}

\begin{figure}
\epsfysize16cm 
\centerline{\epsfbox{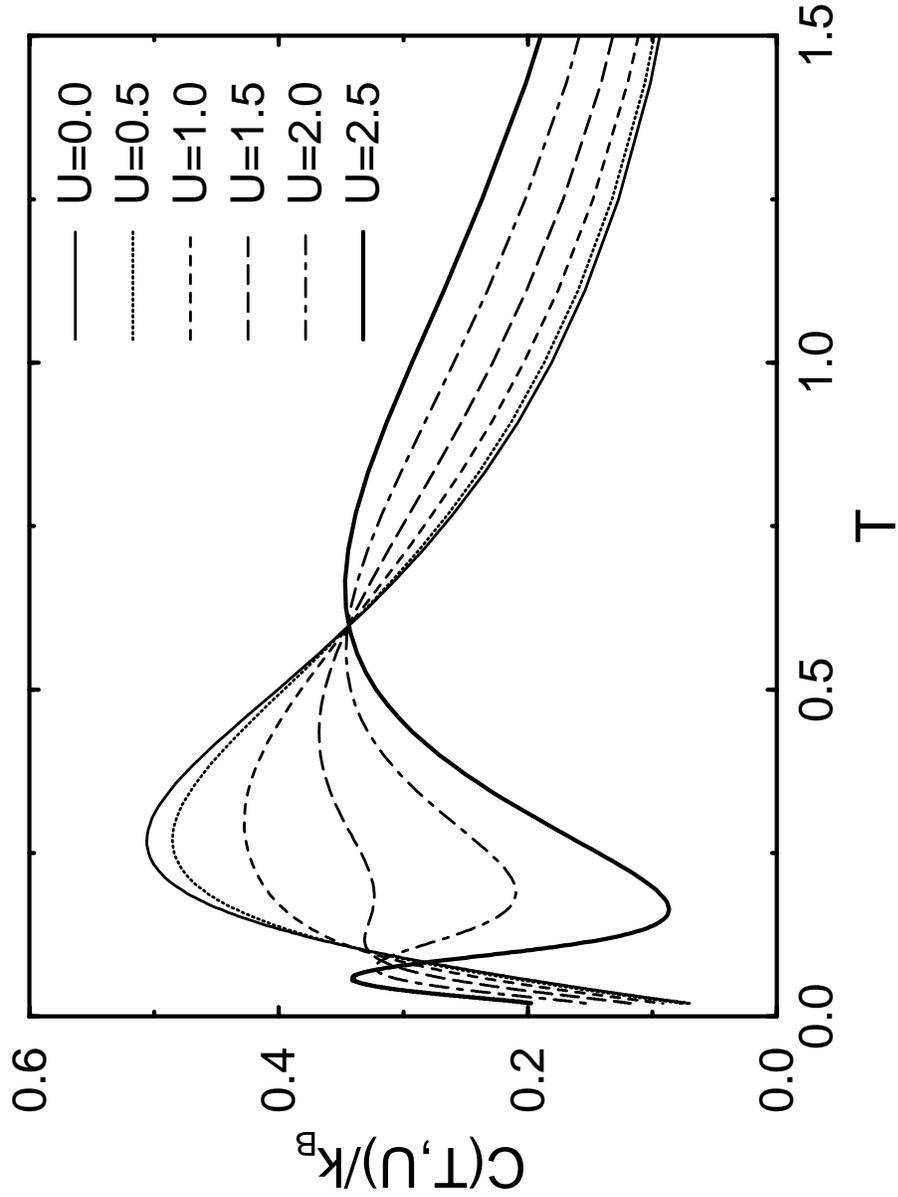}}
\vspace*{2cm}
\caption{Specific heat $C (T,U)$ of the 
paramagnetic phase of the Hubbard model in
$d = \infty$ dimensions calculated by iterated perturbation theory
\protect\cite{5a}.}
\end{figure}

\begin{figure}
\caption{Specific heat (a) $C(T,P)/T$ of CeAl$_3$  
\protect\cite{6}(for $T > 8$K
we took the running average of the data points to reduce the scatter),
(b) $C(T,B)$ of CeCu$_{5.5}$Au$_{0.5}$  \protect\cite{10}.}
\end{figure}

\begin{thebibliography}{99}
\bibitem{17} 
According to Greywall's data \cite{2} Fermi liquid 
theory is valid in $^3$He only below a temperature 
$T_{FL} (P)$, with $T_{FL} (0) \simeq
0.16$K to $T_{FL} (30 {\rm bar}) \simeq 0.1$K.
\bibitem{18}
It is interesting to note that $T_+/T_+^\prime \simeq 4$ both in 
CeAl$_3$ and UBe$_{13}$
and that the values of $C/T$ of the two systems
at $T^\prime_+$ and $T_+$, respectively, 
are also similar.
\bibitem{21}
With $T_+ \simeq 1.1$K and $\partial^2 \chi^{(3)}
(T_+)/\partial T^2 \simeq - 2 \times 10^{-3} [\mu_B/T^3K^2]$, obtained from
Fig.~5a of ref.~\cite{10}, we find $|W(B)| \alt0.1$ for
$B \alt 3 T$ along the c-direction.
\bibitem{23}
The crossing of curves reported in Fig. 4 of ref.\cite{5} has a particularly
small width, 
$|W (U) | \alt 0.004$, caused by an almost perfect
$U^2$-dependence of $E(T,U) - U/2$ up to $U = 2.5t^*$. It is not
clear at present whether this is
partly caused by the 
``iterated perturbation theory''  itself, which iterates
the second-order weak-coupling approximation to the self-energy.
\end{thebibliography}
\end{document}